# Maximizing the Collective Learning Effects in Regional Economic Development

Jian Gao

Big Data Research Center, University of Electronic Science and Technology of China, Chengdu 611731, China.
MIT Media Lab, Massachusetts Institute of Technology, Cambridge, MA 02139, USA.
E-MAIL: gaojian08@hotmail.com

**Abstract:**

Collective learning in economic development has been revealed by recent empirical studies, however, investigations on how to benefit most from its effects remain still lacking. In this paper, we explore the maximization of the collective learning effects using a simple propagation model to study the diversification of industries on real networks built on Brazilian labor data. For the inter-regional learning, we find an optimal strategy that makes a balance between core and periphery industries in the initial activation, considering the core-periphery structure of the industry space--a network representation of the relatedness between industries. For the inter-regional learning, we find an optimal strategy that makes a balance between nearby and distant regions in establishing new spatial connections, considering the spatial structure of the integrated adjacent network that connects all regions. Our findings suggest that the near to by random strategies are likely to make the best use of the collective learning effects in advancing regional economic development practices.

## 1. Introduction

Economic development is the process in which economies learn to develop new industries and product new products, however, our understanding of its underlying mechanisms is still insufficient and facing challenges [1][2]. Fortunately, with new large-scale data produced by complex economic systems [3][4] and novel analytic tools borrowed from interdisciplinary fields [5][6], the recent research paradigm has been able to deal with the emerging complexity in real-world economic systems [7-9]. In particular, literature has revealed the effects of collective learning [10][11]--the learning that takes place at the scales of groups, organizations, regions and nations--in understanding the basic principles that govern economic development.

Collective learning has been studied extensively using different types of data, at different scales, and across different contents. In particular, recent studies have highlighted the collective learning effects in two channels [10][11]. One is the inter-industry learning channel, which focuses on the effects of learning from related economic activities. For example, regions are more likely to diversify into industries that are more related to their current industries [12][13]. The other is the inter-regional learning channel, which focuses on the effects of learning from geographic neighbors. For example, countries have higher probability to export (import) a product if their neighboring countries have already exported (imported) that product [14]. More interestingly, recent empirical works find that the two collective learning channels work as substitutes [10][11].

However, investigations on the best development strategy to benefit most from the two collective learning effects for regions with different preexisting industries is still missing. One promising step is to study the diversification of industries using simulations on real networks by employing spreading models [15][16], where industries or regions are more likely to be activated if they already have more active neighbors. Besides, the role that the structure of the underlying networks plays on advancing or suppressing industrial diversification is not yet fully understood. For inter-industry learning, the industry space--a network representation of the relatedness between industries--has the core-periphery structure [17], resulting in different costs to activate industries at different network locations [18]. For inter-regional learning, one region could connect to distant regions through spatial links (like airlines), which makes the spreading dynamic more complex [19].

In this paper, we study the maximization of the collective learning effects in industrial diversification by doing simulations on real networks using a simple propagation model, in which an industry or a region will be activated if over half of its neighboring industries or regions are already active. For the inter-industry learning, we find that the optimal strategy that makes a balance between core and periphery industries in the initial activation can give a full final activation of all industries within short time. For the inter-regional learning, we find that the optimal strategy that makes a balance between nearby and distant regions in establishing new spatial connections can give a full final activation of all regions with short time and low costs. Further, we discuss the promising applications of these findings to real-word economic development practices.

## 2. Data and Model

We first brief introduce the underlying networks and the propagation model. For the inter-industry learning, the network is the industry space that connects industries. For the inter-regional learning, the network is the adjacent network of regions. For the activation of industries, the model is a simple threshold propagation process.

The industry space is a network representation of the relatedness between industries, which is measured by their co-hiring of occupations based on the Brazilian labor data (RAIS). Specifically, two industries have a higher relatedness if they are more likely to hire for the same occupation. Based on the relatedness matrix, the industry space is built by overlapping the maximum spanning network and the maximum weighted network (see Refs [10][11] for details). Figure 1(a) presents the Brazilian industry space, showing relationship among 669 industries at the Class level with the average degree being at about 6.5. Each node in the industry space has an coreness value, calculated by using the k-shell decomposition [20].

The adjacent network is built based on the geographic neighboring relationship among regions. Two regions are connected by an undirected and unweighted link if they share border. Based on the Brazilian data, the adjacent network presents relationship among 558 Microregions with the average degree being at about 6, as shown in Fig. 1(b). Moreover, the neighboring distance $d$ between two regions is defined as the minimum number of regions that one region has to cross to reach the other region. By definition, $d = 1$ for two neighboring regions.

The model to simulate the activation of industries is a threshold propagation process on networks [15]. In the network: (i) Nodes are in either active or inactive status; (ii) Nodes remain active once activated. For the activation process [19]: (i) A given ratio of nodes ($p$) are initially activated; (ii) Inactive nodes become active if over half of their neighbors are already active; (iii) Inactive nodes are activated in an iterative manner until reaching the steady state. Two metrics of interest are: the relative size of the final active nodes to all nodes ($S_a$), and the number of iterations (time step) to the final activation ($NOI$). Figure 1(c) illustrates the propagation process, where two nodes are initially activated (colored black), i.e., $p = 2/8$. After the propagation, $S_a = 5/8$ and $NOI = 3$.

## 3. Results

### 3.1. Inter-industry learning

For the inter-industry learning, the strategy decides which set of industries are suggested to be initially developed. Considering that the industry space has the core-periphery structure where core-located nodes are highly connected with each other while periphery-located nodes connect a few nodes, different industries not only face different opportunities to be developed but also have different powers to further active other neighboring industries. Therefore, there expected to be an optimal strategy in choosing the initial active industries to maximize the benefits from the inter-industry learning effects.

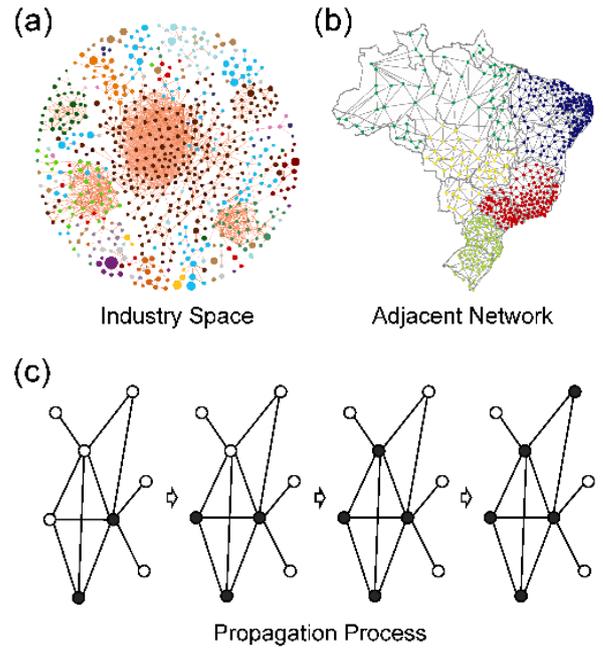

Figure 1. The underlying networks and the propagation process. (a) The network representation of the Brazilian industry space. (b) The Brazilian adjacent network. (c) The illustration of the threshold propagation process.

In the simulations, the ratio of the initially activated industries is set as $p$, but the set of these industries is selected according to the balance index of core and periphery industries ($q$) in the industry space as shown in Figure 2(a). The selection process works as follows. First, a randomized list consisting all industries is generated. Then, $q$ ratio of industries are randomly selected from the list and rearranged by their coreness in the network in descending order (to generate cases that $q$ varies from 0 to 1) or in ascending order (to generate cases that $q$ varies from 0 to -1). Finally, $p$ ratio of top-listed industries in the rearranged list are selected to be initially activated. In short, the balance index $q = -1$ means always selecting periphery-located industries as in Figure 2(b), $q = 0$ means selecting industries by random as in Figure 2(c), and $q = 1$ means always selecting core-located industries as in Figure 2(d).

Figure 3(a) presents the phase diagram where the color corresponds to $S_a$, the horizontal-axis is $p$, and the vertical-axis is the balance index of core and periphery industries $q$. We find that the diagram is trivial when $p < 0.3$ or $p > 0.8$ where different strategies perform almost the same. However, a nontrivial area emerges in the middle of the diagram, where the near to by random strategy (with $q$ being around 0) is more likely to give the full activation of industries at the end ($S_a = 1$).

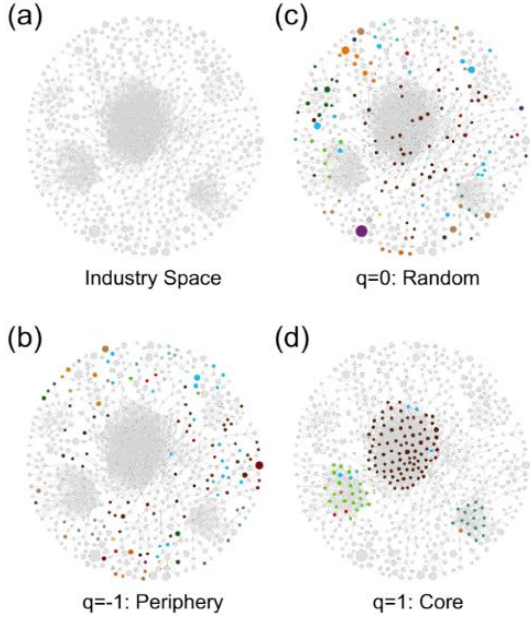

Figure 2. The selection of initially activated industries. (a) The base Brazilian industry space without active industries. (b) The initially activated periphery-located industries with $q=-1$, (c) by random with $q=0$, and (d) core-located industries with $q=1$. For all illustrations, the ratio of the initial activation is set as $p=0.2$. Active nodes are highlighted by their original color, otherwise by gray.

In particular, when the initial ratio $0.3 < p < 0.5$, to initially active core-located ($q=1$) or periphery-located ($q=-1$) industries is not competitive, because only part of industries can be finally activated, as shown in Figure 3(a). By comparison, the by random strategy ($q=0$) performs the best by giving the full activation of all industries and taking short time. When the initial ratio $0.5 < p < 0.8$, to initially active periphery-located industries is the worst strategy, because only about half industries can be finally activated. To initially active core-located industries is the best strategy, because it gives the full final activation as in Figure 3(a) and takes shorter time as in Figure 3(b).

### 3.2. Inter-regional learning

For the inter-regional learning, the strategy decides whether to choose nearby or distant regions to establish new spatial connections. For a region, building new connections will change the density of its neighboring economic activities, leading to different opportunities for its future development. Building rails will easily connect nearby regions with relative low costs, while opening flights can significantly reduce the commuting time between distant regions, making it as if they are neighbors, but with relative large costs. Therefore, there expected to be a nontrivial strategy in determining the length of newly established spatial connections among regions to maximize the benefits from the inter-regional learning effects.

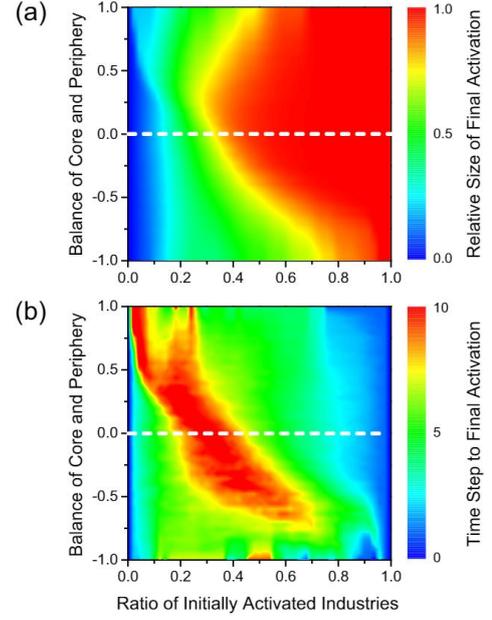

Figure 3. The simulation results for maximizing the inter-industry learning effects. (a) $S_a$, as shown by the color. (b) $NOI$, as shown by the color. The horizontal-axis is $p$, and the vertical-axis is $q$.

In the simulations, the initially activated regions with ratio $p$ are randomly selected, but the integrated adjacent networks are built by adding one new spatial link between each pair of regions in the original adjacent network as in Figure 4(a), where the length of the spatial link is determined by the balance index ($Q$) of nearby and distant regions. The establishment of spatial links works as follows. First, for each region, a random distance $r$ between 2 and $D/2$ is generated with probability $P(r) \propto r^{5Q}$ [19]. The distance $D$ is the maximum neighboring distance between that region and all other regions. The decay parameter $5Q$ is used to approach the boundary conditions, where the length of spatial links is the longest or the shortest. Then, to establish the spatial link, one region is randomly selected from the set of candidate regions with $r$ neighboring distance. Finally, the procedure is repeated to finalize the integrated adjacent network, where each region has an undirected spatial link. In short, the balance index $Q=-1$ means always linking to nearby regions as in Figure 4(b), $Q=0$ means linking to regions by random as in Figure 4(c), and $Q=1$ means always linking to distant regions as in Figure 4(d).

Figure 5(a) presents the phase diagram where the color corresponds to $S_a$, the horizontal-axis is $p$, and the vertical-axis is the balance index of nearby and distant regions ($Q$). When $p < 0.18$ or $p > 0.24$, we find that the diagram is trivial. However, non-trivial diagram emerges at the middle part, showing that the by random strategy ($Q=0$) and the distant-preferred strategy ($Q>0$) give the full activation of all regions as in Figure 5(a).

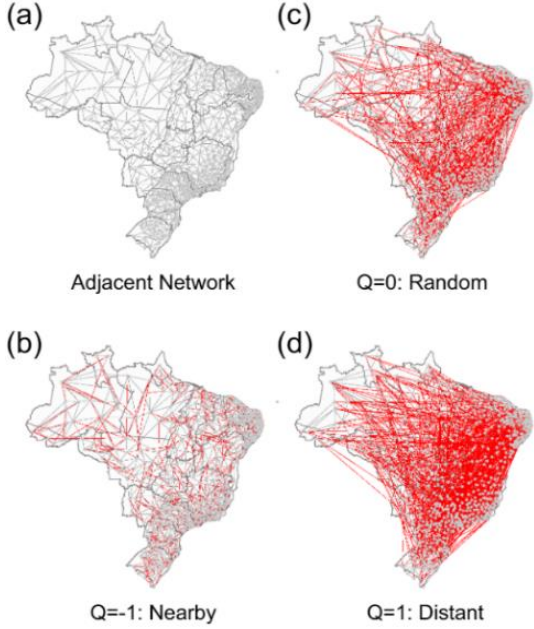

Figure 4. The Brazilian regional adjacent network integrated with new spatial links. (a) The original adjacent network. (b) The integration adjacent network by adding spatial links to nearby regions with $Q=-1$, (c) to regions by random with $Q=0$, and (d) to distant regions with $Q=1$. The neighboring links are in gray, and the spatial links are in red.

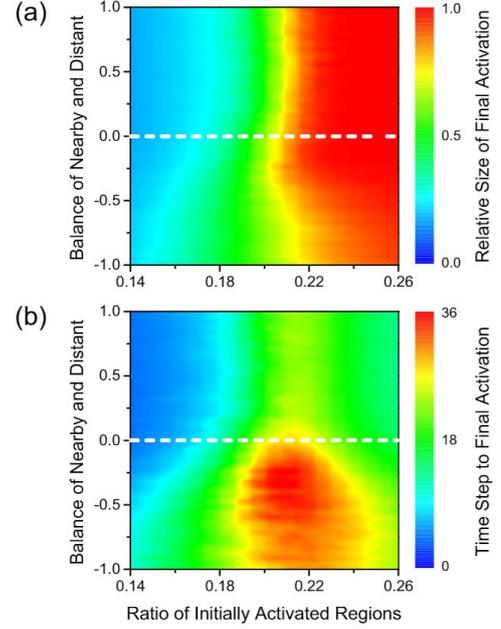

Figure 5. The simulation results for maximizing the inter-regional learning effects. (a) $S_a$, as shown by the color. (b) $NOI$, as shown by the color. The horizontal-axis is $p$, and the vertical-axis is $Q$.

In particular, when the initial ratio $p<0.18$, the nearby-preferred strategy ($Q<0$) not only actives more industries as in Figure 1(a), but also takes more time as in Figure 1(b). When the initial ratio $0.18<p<0.21$, all strategies give partial activation of all regions, but the by random strategy ($Q=0$) and distant-preferred strategy ($Q>0$) are the most efficient ones since they take the shortest time. When the initial ratio $0.21<p<0.24$, the nearby-preferred strategy is the worst one, because it only actives part of all regions but takes the longest time. By comparison, the by random strategy and distant-preferred strategy both give the full activation of all regions. In short, we find that the random connecting strategy (for example, the combination of opening long-distance flights and building short-distance rails) performs as the best as the distant-preferred strategy (only opening flights) but may save construction and operating costs.

## 4. Conclusions and Discussion

In this paper, we explored the maximization of collective learning in regional economic diversification by employing a threshold propagation model to do simulations on real networks. For the inter-industry learning, we proposed the balance index of core and periphery industries in the industry space to control the selection of initially activated industries. We found the near to by random strategy is an optimal strategy in the initial activation. For the inter-regional learning, we proposed the balance index of nearby and distant regions to control the establishment of new spatial connections among regions. We found the near to by random strategy is also an optimal strategy in establishing new spatial connections. These findings shed some light on making the best use of the collective learning effects in regional economic development.

Some challenges in understanding the mechanisms of industrial diversification still remain, and our analysis should be interpreted in light of its inevitable limitations. On the one hand, it will be an improvement if both effects of inter-regional learning and inter-regional learning can be considered at the same time as their interactions are not yet fully understood in facilitating industrial diversification. On the other hand, the robustness could be further tested by employing a variety of models to simulate industrial diversification on networks with different structures.

**Acknowledgements**

Jian Gao acknowledges Cesar A. Hidalgo, Bogang Jun, Flavio Pinheiro, and Tao Zhou for helpful discussions, the Collective Learning Group at the MIT Media Lab for providing the networks and computing resources, and the China Scholarship Council for partial financial support.